\documentclass[preprint,12pt,authoryear]{elsarticle}
\usepackage{amssymb}

\usepackage[process=auto,crop=preview]{pstool}
\usepackage{booktabs}
\usepackage{amsmath}
\DeclareMathOperator\erf{erf}

\journal{International Journal of Multiphase Flow}
\begin{document}

\begin{frontmatter}

\title{Turbulence and self-similarity in
highly-aerated shear flows: the stable hydraulic jump}

\author[mymainaddress1]{M. Kramer
\corref{mycorrespondingauthor}}
\cortext[mycorrespondingauthor]{Corresponding author}
\ead{m.kramer@adfa.edu.au}
\ead[url]{https://www.unsw.adfa.edu.au}
\address[mymainaddress1]{University of New South Wales, School of Engineering and Information Technology (SEIT), Canberra,
ACT 2610, Australia}

\author[mymainaddress2]{D. Valero}
\address[mymainaddress2]{IHE Delft Institute for Water Education,Water Science and Engineering Department, 2611 AX Delft, The Netherlands}

\begin{abstract}

Hydraulic jumps are oftentimes encountered in natural and human-made environments. The transition from supercritical to subcritical flow involves large energy dissipation rates and substantial air entrainment, preventing the use of monophasic flow measurement instrumentation. This paper presents an experimental study of a stable hydraulic jump with a Froude number of 4.25, utilizing novel intrusive phase-detection probe techniques and image-based velocimetry from a side perspective. Turbulence estimations were obtained for the impinging region and the roller region of the jump including  Reynolds stresses, turbulent integral scales and velocity fluctuations spectra. The velocity spectra 
had a $-$5/3 slope in the inertial subrange and flattened at larger frequencies. This is thought to be linked to an energy transfer from the inertial range to the frequencies associated with bubble scales. Overall, the collected data is of particular interest for high-fidelity numerical model validation and the study represents an advancement in air-water flow research.

\end{abstract}


\end{frontmatter}

\section{Introduction}\label{sec:intro}
Hydraulic jumps have been studied for at least two centuries, with the first steps taken by Bidone \citep{Bidone1820, Mossa03} and B\'{e}langer \citep{Belanger1828, Belanger1840, Chanson09Bel} following observations by Leonardo Da Vinci more than 500 years ago \citep{Hager92jump}. This phenomenon is governed by  momentum compatibility between supercritical and subcritical flow and can be  described in terms of the inflow Froude number \mbox{$Fr_1 = U_1 / \sqrt{g \, d_1}$;} with $g$ being the gravity acceleration, $U_1$ the time- and depth-averaged velocity and $d_1$ the mean flow depth in the inflow section.

At the toe of the jump, a high velocity water jet impacts a slower mixture of air and water, creating an inflection point of the mean velocity profile. This singularity represents a sufficient condition for inviscid instability, as it satisfies the Raileigh-Fj{\o}rtof conditions \citep{Drazin02}. The inflow boundary layer interacts with the developing shear layer of the jump and viscous forces become relevant, represented by the Reynolds number $Re = \rho \, U_1 \, d_1 / \mu$; with $\rho$ and $\mu$ representing the water density and the dynamic viscosity, respectively.

\begin{figure}[h!]
\centerline{\includegraphics[width=0.8\textwidth]{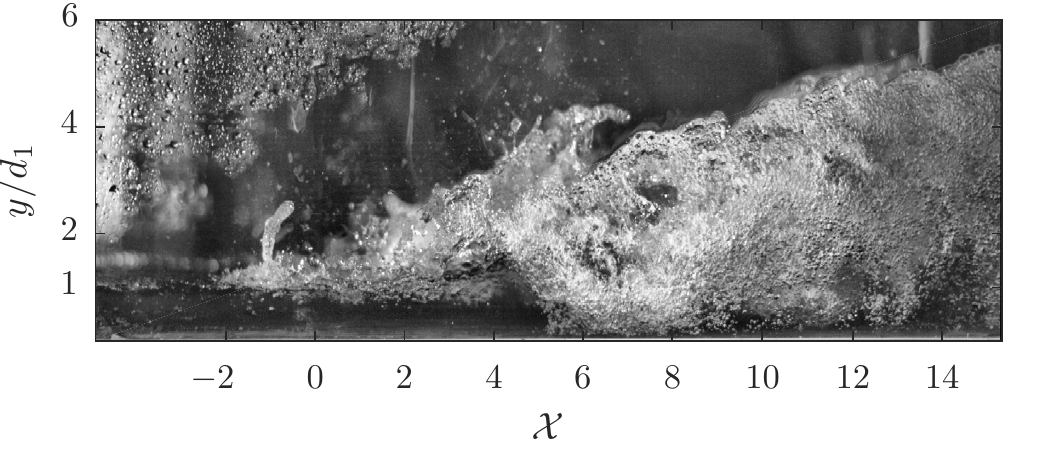}}
\caption{Impinging and roller region of a hydraulic jump; $\mathcal{X}=(x-x_1)/d_1$ with $x$ being the streamwise coordinate and $x_1$ the jump toe location.}
\label{fig:ExperimentalSetup}
\end{figure}

With increasing inertial and turbulent forces, the surface tension is exceeded and considerable volumes of air are entrained \citep{Brocchini01JFM} both at the impingement point and at the free-surface of the roller (figure \ref{fig:ExperimentalSetup}). The ratio of inertial to surface tension forces can be described in terms of the Weber number $We = \rho \, U_1^2 d_1 / \sigma$; with $\sigma$ being the air-water surface tension.

Given that the entrainment and transport of air prevents the use of classic measurement instrumentation, instantaneous velocities and turbulence parameters have remained poorly understood in the aerated region of hydraulic jumps. Some studies have addressed turbulence quantities for the water phase \citep{Rouse58} and for jumps with low air concentrations \citep{Long1990LDA, Liu04, Lennon06}, typically at Froude numbers $Fr_1 <3$. In highly-aerated jumps, the mean flow structure has been analysed by several researchers using Prandtl-type pitot tubes and phase-detection intrusive probes \citep{Rajaratnam65, Chanson20, Murzyn05}. Free-surface fluctuations were obtained with electric hydrometers, acoustic displacement meters (ADM) and recently using LIDAR technique \citep{Mossa99osc,Wang2015, Montano18}. Arrays of wire gauges and phase-detection probes were used to characterise interfacial correlation length scales and transverse motion \citep{Mouaze05,Wang2017,Wang19}.

\cite{Mossa98} drew attention to the potential use of imaging techniques for the study of hydraulic jumps and recent developments in this area have now enabled the estimation of velocity fluctuations in highly-aerated flows \citep{Lin12,Bung16OF,Zhang17,Kramer18OF}. 
Similarly, pseudo-instantaneous velocity measurements with dual-tip phase-detection probes were recently obtained by \cite{Kramer19AWCC}, while phase-detection probes have been previously used to study mean velocities only \citep{Chanson20,Murzyn07,Chanson09,Wang14}. By allowing the estimation of instantaneous velocities, these newly developed intrusive and imaging approaches overcome the limitations of previous experimental studies, thereby enabling cross-comparison between independent measuring instruments.

Numerical modelling has provided researchers with an alternative tool to gain further insights into the  flow structure of hydraulic jumps. The direct numerical simulation (DNS) of \cite{Mortazavi16DNS} compared mean flow variables and scales of free-surface fluctuations to previous experimental literature, with an overall good agreement. \cite{Jesudhas16DES} demonstrated the evolution of the shear layer and identified coherent structures using detached eddy simulation (DES). \cite{Witt15} and \cite{Bayon16} discussed the flow structure of hydraulic jumps and showed good agreement of mean flow variables between eddy viscosity models and experimental observations.

Nevertheless, validation data for the most relevant turbulent quantities in the aerated region of hydraulic jumps are rare. This study explores the turbulent structure of a stable hydraulic jump \citep[$Fr_1 >$ 4,][]{Montes98} using novel experimental techniques, enabling turbulence estimations in most energetic and highly aerated flow regions. 

\section{Experiment}\label{sec:Experiment}
\label{sec:exp_setup}
\subsection{The water channel}
Laboratory experiments were performed in a horizontal, rectangular channel with a length of 3.2 m, a width of $W$ = 0.5 m  and a height of 0.4 m. The channel had a smooth PVC bed, glass sidewalls, a head tank and an upstream vertical undershoot gate with a semi-circular rounded shape. The head tank was equipped with a series of flow straighteners to ensure smooth inflow conditions. During the experiments, the opening of the undershoot gate was set to 0.036 m.

The jump toe was located at a distance of \mbox{$x_1$ = 1.4 m} from the upstream gate (partially developed inflow conditions), and the inflow depth and velocity were \mbox{$d_1$ = 0.042 m} and \mbox{$U_1$ = 2.73 m/s}, corresponding to Froude, Reynolds and Weber numbers of \mbox{$Fr_1$ = 4.25}, $Re$ = 1.15$ \times $10$^5$ and $We$ = 3.93$\times $10$^3$, respectively.

The produced hydraulic jump was controlled with a vertical overshoot gate in the tailrace of the channel, yielding a flow depth of $d_2$ = 0.24 m at the end of the jump. Further details on the experimental channel can be found in \cite{Wang14}.

\subsection{The instrumentation set-up}
The deployed instrumentation included intrusive and non-intrusive measurement techniques. The flow rate was estimated by means of a Venturi meter installed within the feeding pipe of the head tank, with an expected accuracy of $\pm$2$\%$. Free-surface elevations upstream and downstream of the jump were measured with a pointer-gauge.

Image sequences of the highly-aerated hydraulic jump flow were filmed using a Phantom v2011 high-speed video camera, placed at a distance of 2 m to the sidewall and equipped with a Zeiss 85 mm planar lens, holding a small distortion level. The scene was illuminated with a 4$\times$6 high power LED lamp (GS Vitec MultiLED). The camera was focused on the flow next to the sidewall (figure \ref{fig:ExperimentalSetup}) and the depth of field (DOF) approximately spanned within 3 mm from the inside wall. Video signals were recorded for a sampling duration of 10 s and a sampling rate of 5 kHz with HD resolution (\mbox{1280$\times$512 px$^2$}).
Note that one pixel on the image plane corresponded to a physical length of 0.66 mm, which was equal to a pixel density of 15 px/cm.

A dual-tip phase-detection conductivity probe (inner diameter: 0.25 mm, outer diameter: 0.8 mm, streamwise tip separation: 4.7 mm)  was used to measure instantaneous air-water flow properties at the centreline of the channel ($2z/W = 0$) and next to the sidewall ($2z/W = 0.96$), intersecting the camera's DOF. Measured air-water flow properties included void fraction $C$, bubble/droplet count rate $F$ and streamwise interfacial velocities $U$. The probe was mounted on a trolley system and its position was monitored with a digimatic scale unit, having a positioning accuracy of $\pm$0.025 mm. The sampling rate and duration were 20 kHz and \mbox{90 s} respectively.

The probe was reversed in the roller region. When the probe is aligned with the flow streamlines, the accuracy of the recorded time-averaged interfacial velocities $\langle U \rangle$ can be considered to be within $\Delta\langle U \rangle$/$\langle U \rangle<5\%$ for \mbox{$0.05<\langle C \rangle<0.95$} and $\Delta\langle U \rangle$/$\langle U \rangle<$ 10$\%$ for $0.01<\langle C \rangle<0.05$ and $0.95<\langle C \rangle <0.99$ \citep{Carosi08} (the operator $\langle... \rangle$ denotes time-averaging). In highly three-dimensional flow regions, the accuracy is known to be lower than in regions with streamline alignment of the tips \citep{Kramer19AWCC}.

\subsection{The velocity estimation method}
The video signals were analysed with the Farnebaeck method (\ref{Farnebackmethod}). This image-based velocimetry technique belongs to the class of optical flow (OF) methods and has recently attracted interest in the study of air-water flows, proving to produce less errors in bubbly flow images \citep{Bung16OF,Zhang17,Kramer18OF}.

Phase-detection conductivity probe (CP) signals were processed with the adaptive window cross-correlation (AWCC) technique (\ref{AWCCT}), allowing for the estimation of pseudo-instantaneous and near-null velocities \citep{Kramer19AWCC,AWCC}.

\section{Results and discussion}\label{sec:results}
\subsection{Conjugate depths, jump length and free-surface profile}
\begin{figure}[h!]
\centerline{\includegraphics[width=0.95\textwidth]{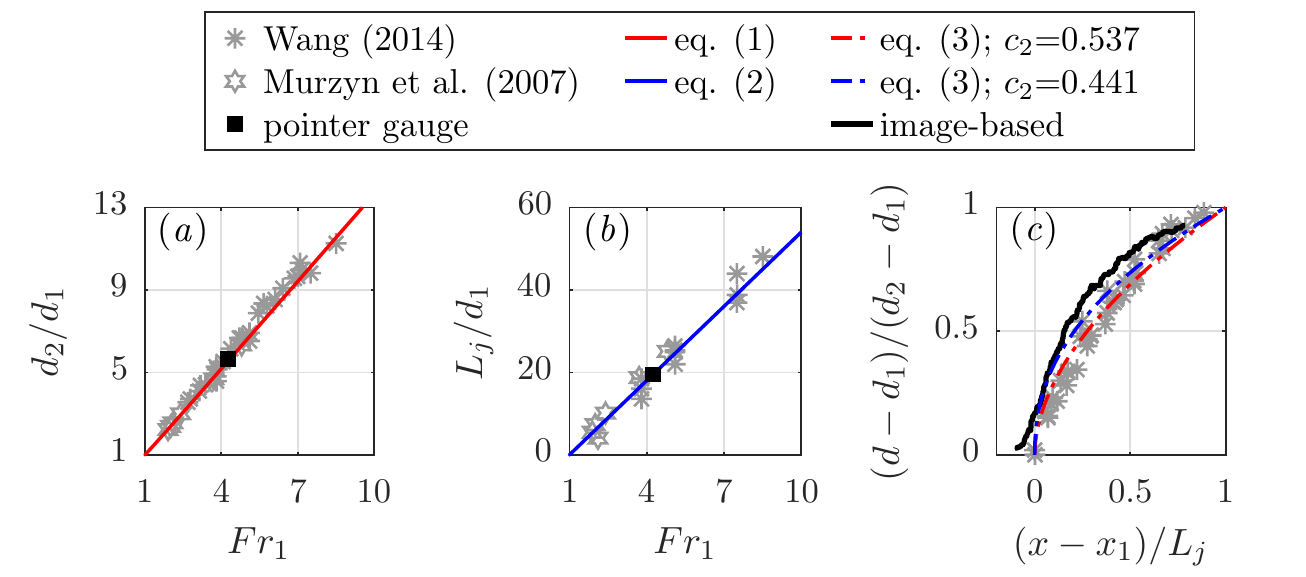}}
\caption{Basic hydraulic jump properties -- comparison of pointer gauge and OF measurements with data from \cite{Wang14} and \cite{Murzyn07a}
\protect (\textit{a}) conjugate depth ratio as function of the inflow Froude number \protect (\textit{b}) jump length $L_j$ as function of $Fr_1$ \protect (\textit{c}) self-similar free-surface profile with $d$ = free-surface elevation.}
\label{fig:freesurface}
\end{figure}

The relationship between upstream and downstream flow depths in hydraulic jumps can be described by the well-known \cite{Belanger1840} momentum equation, which was derived under the assumption of negligible boundary shear stress, uniform velocities and hydrostatic pressure distributions at the beginning and end of the jump:

\begin{equation}
\label{eq:belanger}
\frac{d_2}{d_1}=\frac{1}{2} \Big( \sqrt{1+8 Fr_1^2}-1\Big)
\end{equation}
where the subscripts 1 and 2 refer to upstream and downstream flow conditions. Figure \ref{fig:freesurface} (\textit{a}) shows  eq. (\ref{eq:belanger}) together with conjugate depths of the current investigation and data from \cite{Wang14}. The herein measured conjugate depths differed less than 1 \% from eq. (\ref{eq:belanger}). 

The upstream flow depth $d_1$ is a scaling parameter in the vertical direction, while longitudinal scaling  invokes the jump length ($L_j$). 
The jump length $L_j$ can be defined as the distance between jump toe and the location where the free-surface becomes horizontal \citep{Hager90}. In the studied case, the jump length was \mbox{$L_j$ = 0.82 m} (measured with pointer-gauge, figure \ref{fig:freesurface} (\textit{b})) and was well-represented by the following expression:
\begin{equation}
\label{eq:jumproller}
L_j/d_1 = c_1(Fr_1-1)
\end{equation}
where $c_1$ is an empirical coefficient,
taken as $c_1$ = 6 \citep{Wang14}. Eq. (\ref{eq:jumproller}) can be used to determine the jump length $L_j$ (and similarily the roller length $L_r$) for hydraulic jumps on smooth and rough beds, however different values of $c_1$ must be considered, see \cite{Carollo07,Carollo12,Wang14}. Note that jump length and roller length are not identical and their definition varies across literature.  

The geometry of the time-averaged free-surface of hydraulic jumps was previously discussed, amongst others, by \cite{Bakhmeteff36,Rajaratnam65,Hager93,Chanson11,Wang14}. Based on the length of the jump and considering conjugate depths, one self-similar relationship was found:
\begin{equation}
\label{eq:freesurface}
\frac{d - d_1}{d_2 - d_1} = \Bigg(\frac{x-x_1}{L_j}\Bigg)^{c_2}
\end{equation}
where $d$ is the elevation of the free-surface and the exponent is $c_2$ = 0.441  \citep{Chanson11} and $c_2$ = 0.537 \citep{Wang14}. In the present study, the free-surface elevation was determined via image-based analysis. For each recorded frame, the air-water interface level was identified through an image gradient filter, similar to the indicator function presented in \cite{Kramer18OF}. The time averaged free-surface profile is shown in figure \ref{fig:freesurface} (\textit{c}), together with eq. (\ref{eq:freesurface}) and acoustic displacement meter (ADM) measurements from \cite{Wang14}. In the region $0.25 <(x-x_1)/L_j <0.6$, the image-based measurements were slightly above the ADM data, indicating some sidewall effects. Overall, a good agreement between image-based analysis, ADM and empirical fit was observed.

\subsection{Void fraction and bubble/droplet count rate distributions}
\begin{figure}[h!]
\centerline{\includegraphics[width=0.85\textwidth]{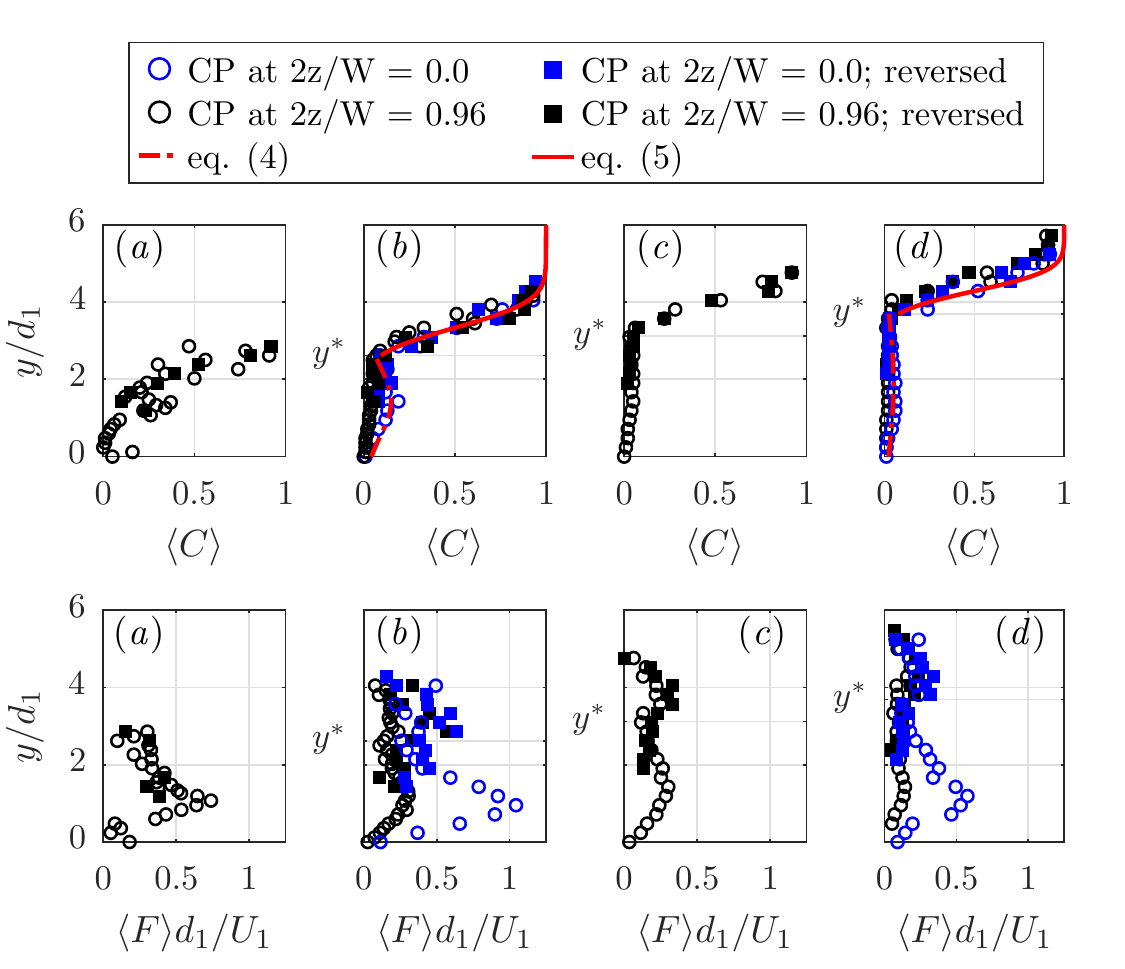}}
  \caption{Void fraction and bubble count rate at different streamwise positions - comparison of phase-detection data recorded at the centreline and near the sidewall; upper row: void fraction $\langle C \rangle$; lower row: dimensionless bubble/droplet count rate $\langle F \rangle d_1/U_1$ (\textit{a}) $\mathcal{X}$ = 3.6 \protect (\textit{b}) $\mathcal{X}$ = 7.1 \protect (\textit{c}) $\mathcal{X}$ = 10.7 and (\textit{d}) \protect $\mathcal{X}$ = 14.3.}
  \label{fig:phase-detection}
\end{figure}

Mean void fraction $\langle C \rangle$ and mean bubble/droplet count rate $\langle F \rangle$ were measured with the phase-detection probe at the centreline ($2z/W = 0$) and next to the sidewall of the channel ($2z/W = 0.96$). Figure \ref{fig:phase-detection} shows the evolution of these parameters in the streamwise direction.

Two main flow regions, comprising a turbulent shear region and a recirculation region could be distinguished, separated by the characteristic elevation $y^*$ where the void fraction had a local minimum (figure \ref{fig:phase-detection}). The void fraction distribution in the shear layer was governend by point source air entrainment at the jump toe, while the distribution in the upper region was due to free-surface aeration. The void fraction distributions followed solutions of the advective diffusion equation for air in water (figure \ref{fig:phase-detection}, upper row) \citep{Chanson89,Chanson20}:

\begin{equation}
\label{eq:Clow}
\langle C \rangle =  C_\mathrm{max}   \exp \left( - \frac{1}{4  D^{\#}} \, \frac{\left(\frac{y - y_{C_\mathrm{max}}}{d_1}  \right)^2}{\frac{x - x_1}{d_1}}  \right) \; \; \mathrm{for} \; \; y < y^*
\end{equation}
\begin{equation}
\label{eq:Cup}
\langle C \rangle = \frac{1}{2} \,  \left( 1 + \mathrm{erf} \left( \frac{y - y_{50}}{2  \sqrt{\frac{D^{*}(x - x_1)}{U_1}}}  \right) \right) \mathrm{for} \; \; y > y^*
\end{equation}
with $C_\mathrm{max}$ being the maximum void fraction in the shear region, $y_{ C_\mathrm{max} }$ the corresponding elevation, $y_{50}$ the elevation where $\langle C \rangle$ = 0.5. The dimensionless turbulent diffusivities of the shear region ($D^{\#}$) and the recirculation region ($D^{*}$) were approximated for the channel's centreline \citep{Wang14}: 

\begin{equation}
D^* = 0.008 \times \exp \left( -3.3 \times \frac{x-x_1}{L_j} \right)
\end{equation}

\begin{equation}
D^{\#} = 0.1 \times \left( 1 - \exp \left(-2.3 \times \frac{x-x_1}{L_j} \right)           \right)
\end{equation}

The bubble/droplet count rate was zero at the channel bed and increased with further vertical distance, reaching a local maximum in the turbulent shear layer (figure \ref{fig:phase-detection}, lower row). A local minimum of the bubble/droplet count rate was seen in the upper part of the shear region, roughly corresponding to $y^*$, followed by a second maximum in the recirculation region. Some sidewall effects were observed in terms of void fraction and bubble/droplet count rate, and both flow variables had lower values next to the sidewall when compared to the centreline. 

In the upper flow region, different probe orientations were tested, including 1.) orientation of the tips against the wall-jet flow direction and 2.) inversed probe. Results were nearly identical, suggesting the presence of highly three-dimensional flow structures. Overall, mean void fraction and bubble count rate were in good agreement with previous measurements \citep{Chanson89,Chanson20,Wang14} and the void fraction distribution followed semi-empirical solutions of the advective diffusion equation.

\subsection{Velocity decay, spreading rate and mean velocity profiles}
\begin{figure}[h!]
\centerline{\includegraphics[width=0.95\textwidth]{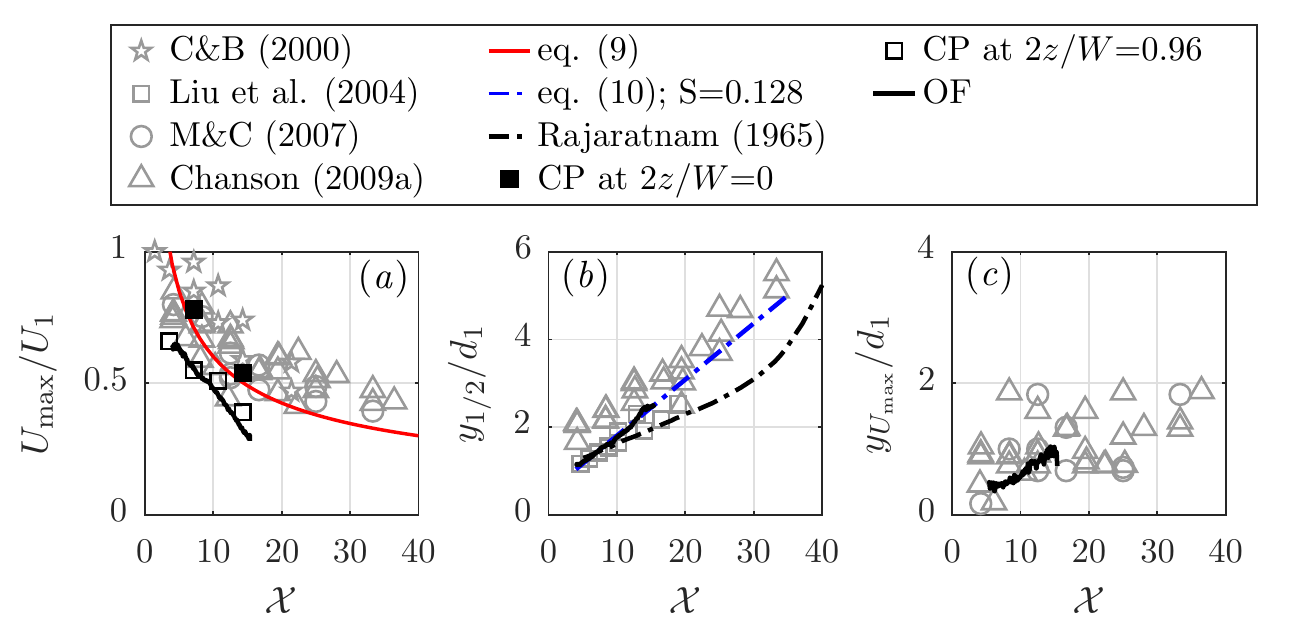}}
\caption{Velocity decay and spread rate of the hydraulic jump - comparison of OF and CP measurements (present study) with data from \cite{Rajaratnam65}, \cite{Chanson20}, \cite{Liu04}, \cite{Murzyn07} and \cite{Chanson09} (\textit{a}) streamwise velocity decay \protect (\textit{b}) half-width of the hydraulic jump (\textit{c}) distance from channel bed where $\langle U \rangle$ = $U _\mathrm{max}$.}
\label{fig:jetdecay}
\end{figure}

Two main features of the impinging and roller regions are the 1.) velocity decay with increasing streamwise distance and the 2.) spreading of the hydraulic jump in the vertical direction. Mean flow velocities were computed using ensemble avaveraging. For the  streamwise component $\langle U \rangle$: 
\begin{equation}
\langle U \rangle = \frac{1}{T}\int_{0}^{T} U(t) \ dt
\end{equation}
where $T$ is the sampling duration and $U$ is the instantaneous flow velocity. The decay of the streamwise velocity of the present study was measured at the channel's centreline (using CP) and near the sidewall (using OF and CP), hinting at some sidewall effects in the order of \mbox{10 \%} for the maximum velocity decay (figure \ref{fig:jetdecay} (\textit{a})). Similar to findings in turbulent plane wall jets \citep{Pope20}, the velocity decayed proportionally to $U_\mathrm{max} \propto x^{-1/2}$, confirming the following empirical relationship (figure \ref{fig:jetdecay} (\textit{a})):
\begin{equation}
\frac{ U_\mathrm{max} }{U_1} = 1.9 \, \mathcal{X}^{-1/2}
\label{veldecay}
\end{equation}
where $ U_\mathrm{max}$ is the maximum cross-sectional velocity and $\mathcal{X} = (x-x_1)/d_1$.

The shear layer of the hydraulic jump can be characterised by means of the spreading rate $S$, traditionally defined as the gradient of the hydraulic jump's half-width $y_{1/2}$ in the streamwise direction
\begin{equation}
S = \frac{\mathrm{d}y_{1/2}}{\mathrm{d}x} 
\label{spreadingrate}
\end{equation}
where the half-width fulfills the following condition
\begin{equation}
\langle U(x,y_{1/2}) \rangle = \frac{1}{2}  U_\mathrm{max} 
\end{equation}
It was found that the investigated hydraulic jump spread linearly (figure \ref{fig:jetdecay} (\textit{b})) at a rate of  $S$ = 0.128, which was in good agreement with data from \cite{Liu04} and \cite{Chanson09}, whereas the spreading rate of \cite{Rajaratnam65} was not constant. A best fit line with $S$ = 0.128 was added to figure \ref{fig:jetdecay} (\textit{b}). Similarily, the elevation where $\langle U \rangle$ = $ U_\mathrm{max}$  showed a linear trend with increasing distance from the jump toe (figure \ref{fig:jetdecay} (\textit{c})), in accordance with previous measurements from \cite{Murzyn07,Chanson09}.

Table \ref{tab:spreadingrate} presents spreading rates for different types of shear flows, indicating that the hydraulic jump spreads at a higher rate than the plane wall jet. This is linked to the adverse pressure gradient together with the multiphase nature of the flow. It must be noted that results of the OF method (\ref{Farnebackmethod}) agreed favourably with previous centreline measurements, suggesting that the mean flow structure remained similar despite the wall proximity (figure \ref{fig:jetdecay}).   

\begin{table}[h!]
\caption{Spreading rate of the classical hydraulic jump compared to wall-jets}
\label{tab:spreadingrate}
\begin{scriptsize}
\begin{center}
\begin{tabular}{l c c c}
\toprule
Reference & flow type & $S$ & $y_{U_\mathrm{max} }/y_{1/2}$   \\
\midrule
\cite{Launder81} & wall-jet & 0.07 $\pm$ 0.01 & 0.13 - 0.17\\
\cite{Rajaratnam90} & wall-jet & 0.068 & -\\
\cite{Eriksson98}& wall-jet & 0.078 & -\\
\cite{Tachie2000} & wall-jet (smooth surface) & 0.085 - 0.090 & 0.15 - 0.20\\
\cite{Tachie2000} & wall-jet (rough surface) & 0.085 - 0.090 & 0.21 - 0.31\\
\midrule
\cite{Rajaratnam65} & hydraulic jump & - & 0.18\\
\cite{Hager92jump} & hydraulic jump & 0.067 & - \\
\cite{Chanson20} & hydraulic jump & 0.109 & 0.25 \\
Present study & hydraulic jump & 0.128  & 0.35 \\
\bottomrule
\end{tabular}
\end{center}
\end{scriptsize}
\end{table}

\begin{figure}[h!]
\centerline{\includegraphics[width=0.95\textwidth]{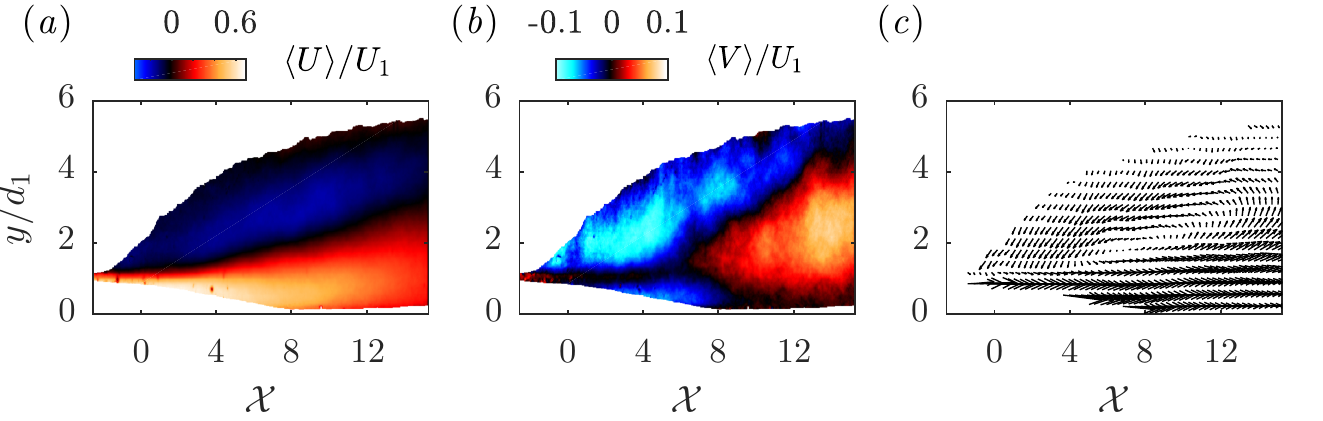}}
  \caption{Mean velocity fields of the hydraulic jump (\textit{a}) streamwise velocity \protect (\textit{b}) vertical velocity (\textit{c}) velocity vector plot.}
\label{fig:OFVelocities}
\end{figure}

The OF method allowed extraction of air-water velocities within the aerated regions, whereas no information was gathered at very low void concentrations; for instance upstream of the jump toe or outside the air-water shear layer (figure \ref{fig:OFVelocities}). The mean streamwise velocity $\langle U \rangle$, dimensionless using $U_1$, showed a deceleration of the impinging water jet and a recirculation region next to the free-surface (figure \ref{fig:OFVelocities}(\textit{a})). Mean vertical velocities $\langle V \rangle$ were smaller than streamwise velocities and included positive and negative values, hence forming the roller of the jump (figures \ref{fig:OFVelocities}(\textit{b,c})). Based on visual observations and computed velocity profiles, the following regions (at $y=d_1$) were identified:
1.) Impinging region ($\mathcal{X}<1$), defined by jump toe fluctuations, splashing and vortex shedding. 2.) Flow establishment zone ($1<\mathcal{X}<5$) with almost horizontal wall-jet. In this region, vortices were advected from the impinging region towards the roller. 3.) Roller region ($\mathcal{X} > 5$) with significant vertical velocities, characterised by smaller bubble sizes and flow recirculation. 

\begin{figure}[h!]
\centerline{\includegraphics[width=0.8\textwidth]{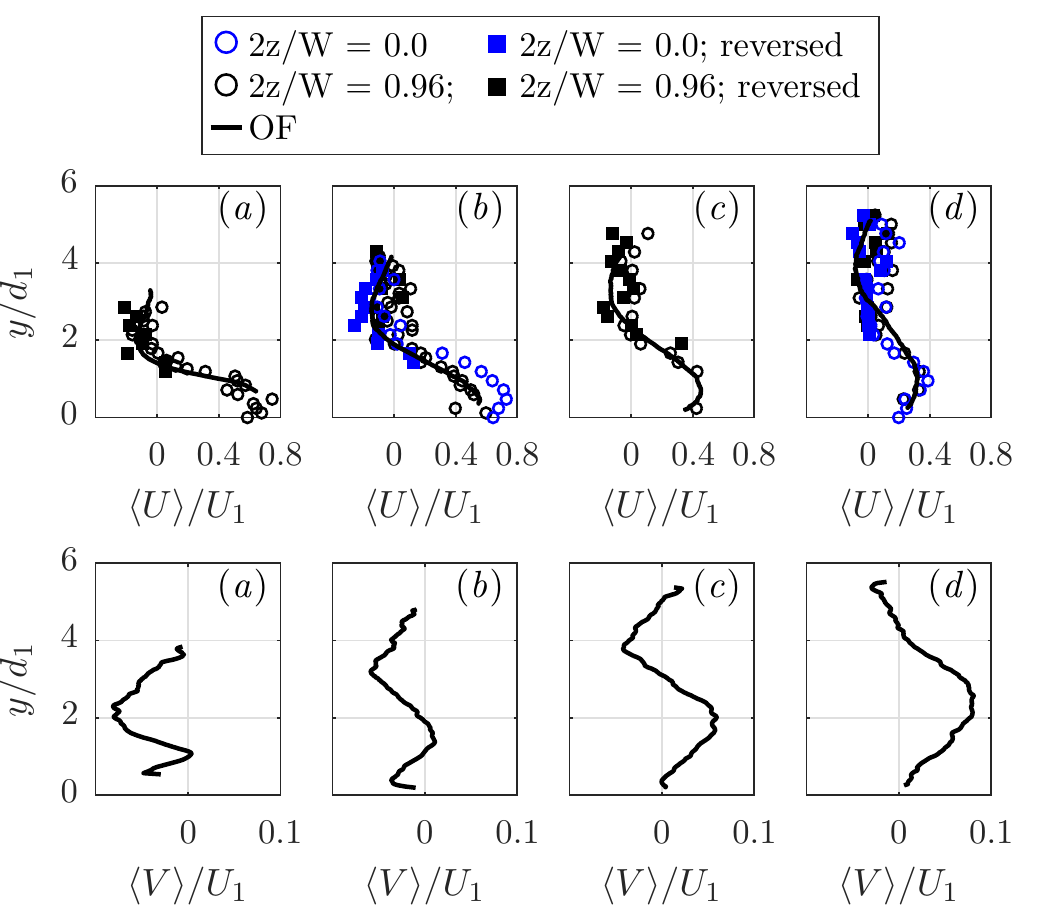}}
\caption{Mean velocity profiles for different streamwise positions, OF compared to CP data near the sidewall ($2z/W=0.96$) and at the centre of the flume ($2z/W=0.0$); upper row: streamwise velocity $\langle U \rangle /U_1$; lower row: vertical velocity $\langle V \rangle /U_1$ (\textit{a}) $\mathcal{X}$ = 3.6 \protect (\textit{b}) \mbox{$\mathcal{X}$ = 7.1} \protect (\textit{c}) $\mathcal{X}$ = 10.7 and (\textit{d}) \protect $\mathcal{X}$ = 14.3.}
\label{fig:VelocityProfiles}
\end{figure}

Figure \ref{fig:VelocityProfiles} shows mean velocity profiles at four positions with different distances from the toe and compares OF (sidewall) and CP data (sidewall, black markers), revealing an excellent agreement of the two independent measurement techniques and therefore endorsing the proposed methodology (figure \ref{fig:VelocityProfiles}, upper row). Note that sidewall effects, as quantified per centreline measurements (blue markers), were in the order of 10 $\%$. The vertical mean velocity profiles, obtained with the OF technique, had positive and negative peaks, reaching up to 10 $\%$ of the inflow velocity, and these were separated by the core of the roller (figure \ref{fig:VelocityProfiles}, lower row).

Besides other shear flows, the hydraulic jump showed a self-similar behaviour, demonstrated by introducing self-similar variables $\zeta$ and $\eta$ in the vertical direction:

\begin{equation}
\zeta = y/y_{1/2}; \quad \eta = \frac{y-y_{ U_\mathrm{max}}}{y_{1/2}-y_{ U_\mathrm{max}}}
\end{equation}
with $y_{ U_\mathrm{max}}$ being the elevation where $\langle U \rangle =  U_\mathrm{max}$ and $y$ the vertical direction. The velocity profiles can be expressed as a function of the self-similar variables
\begin{equation}
f(\zeta) = \frac{\langle U \rangle}{U_\mathrm{max}}; \quad g(\eta) = \frac{\langle U \rangle -  U_\mathrm{min} }{U_\mathrm{max} -  U_\mathrm{min}}; \quad \hat{f}(\zeta) = \frac{\langle V \rangle - V_\mathrm{min}}{ V_\mathrm{max}  -  V_\mathrm{min} }  
\end{equation}
where $f(\zeta)$ and $g(\eta)$ represent the streamwise velocity distribution and $\hat{f}(\zeta)$ the normal velocity distribution. 

\begin{figure}[h!]
\centerline{\includegraphics[width=\textwidth]{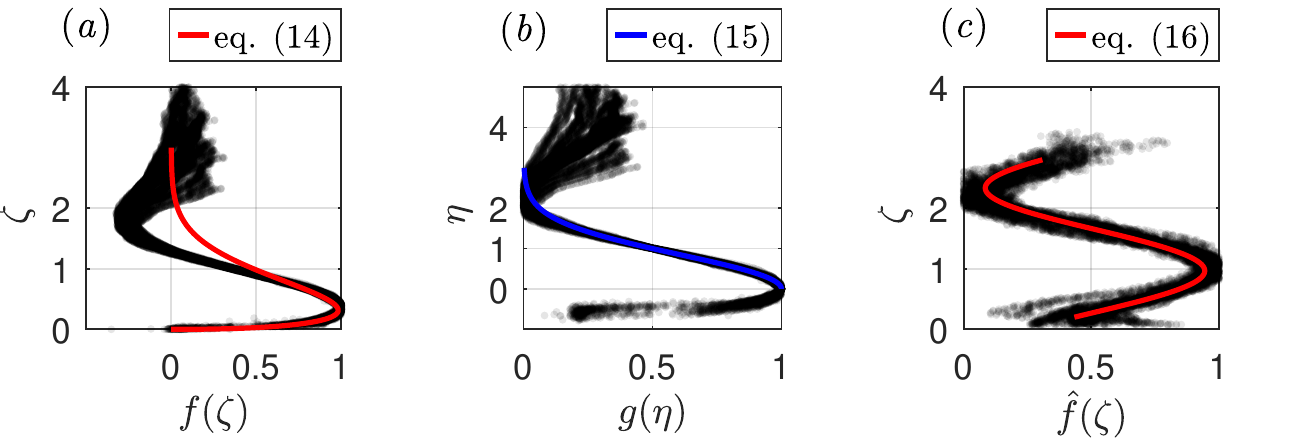}}
\caption{Self-similarity of mean velocity profiles downstream of the impinging region for $\mathcal{X} > 8 $ (\textit{a}) streamwise velocity with variables $\zeta$ and $f(\zeta$) (\textit{b}) streamwise velocity with variables $\eta$ and $g(\eta$) (\textit{c}) vertical velocity  with variables $\zeta$ and $\hat{f}(\zeta$).}
\label{fig:self-similarity}
\end{figure}

As proposed by \cite{Rajaratnam65}, the analogy with a plane wall-jet in the region next to the wall holds. The streamwise velocity of the present investigation followed an empirical wall-jet solution (figure \ref{fig:self-similarity} (\textit{a})), found by \cite{Verhoff64} and adapted by \cite{Lin12}:
\begin{equation}
\label{eq:wall-jet}
f(\zeta) = 2.3 \, (\zeta)^{0.42} \, \left(1-\erf(0.886 \, \zeta) \right)
\end{equation}
where $\erf$ is the error function. Within the upper region, herein defined as $y>y_{1/2}$, the streamwise velocity differed from equation (\ref{eq:wall-jet}) due to the backward flow of the hydraulic jump and $g(\eta)$ was used to represent the self-similar velocity distribution for $\eta > 0$ (figure \ref{fig:self-similarity} (\textit{b})):
\begin{equation}
\label{eq:plane-jet}
g(\eta) = \exp \left( -\alpha \eta^2 \right)
\end{equation}
where $\alpha$ $\approx \ln2$. Note that eq. (\ref{eq:plane-jet}) is commonly used for round and plane free-jets, assuming that the streamwise velocity follows a Gaussian \citep{Agrawal03}. Herein, the parameter $\alpha$ accounted for the backward flow of the hydraulic jump. Figure \ref{fig:self-similarity} (\textit{c}) shows the vertical velocity distribution of the hydraulic jump and it can be seen that 1.) the elevation $y_{ V_\mathrm{max} }$ agreed favourably with the jump's half width, hence $\hat{f}(1) \approx 1$ and $y_{ V_\mathrm{max}} \approx y_{1/2}$, and 2.) the vertical velocity exhibited an S-shaped profile, which could be approximated by a Fourier series:
\begin{equation}
\label{eq:vert-vel}
\hat{f}(\zeta) = \alpha_0+\alpha_1 \, \cos(\zeta \beta) + \alpha_2 \, \sin(\zeta \beta)
\end{equation}
where $\alpha_0 = 0.5144$, $\alpha_1 = -0.2596$, $\alpha_2 = 0.3427$ and $\beta = 2.297$.

\subsection{Velocity fluctuations and Reynolds stresses}
Velocities were decomposed into mean and fluctuating parts as:
\begin{equation}
u = U - \langle U \rangle
\end{equation}
where $u$ is the velocity fluctuation term. The root mean square of velocity fluctuations was computed as:
\begin{equation}
u_{rms} = \sqrt{\langle u^2 \rangle}
\end{equation}
and similarly for the vertical velocity fluctuation $v_{rms}$.

\begin{figure}[h!]
\centerline{\includegraphics[width=0.8\textwidth]{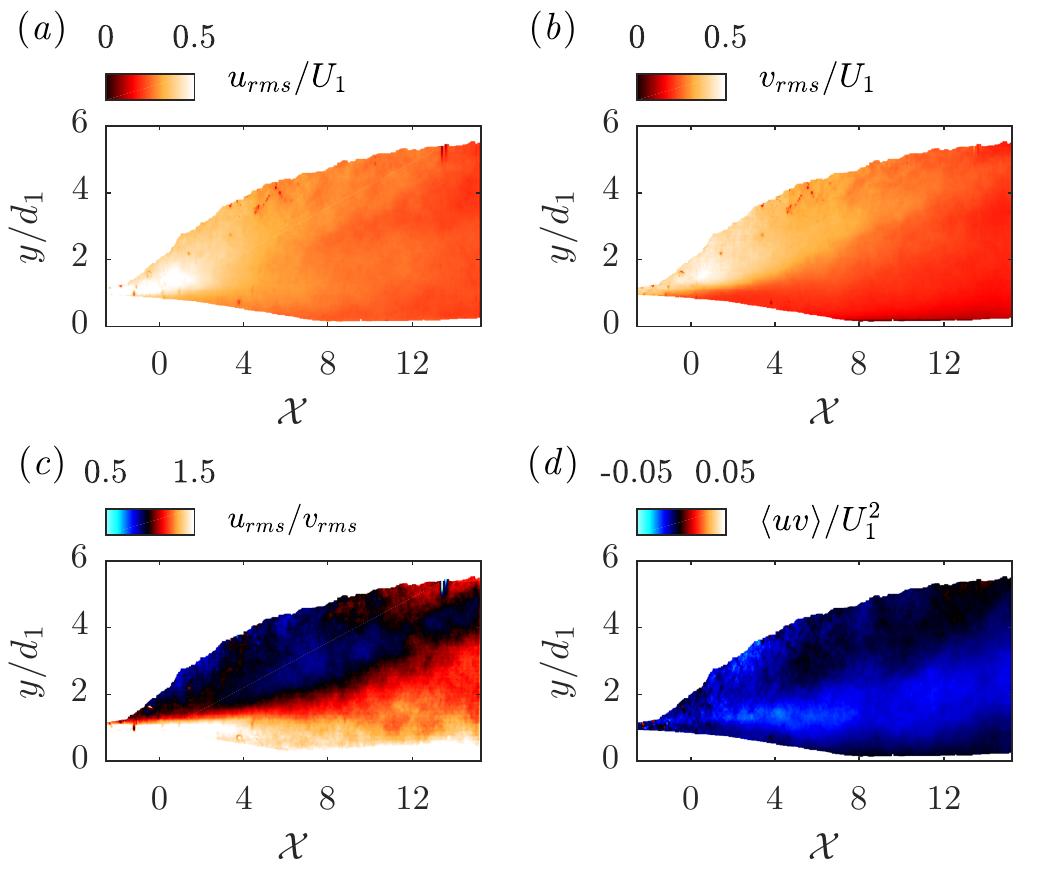}}
\caption{Velocity fluctuations obtained with the OF technique: (\textit{a}) streamwise fluctuations \protect (\textit{b}) vertical fluctuations  (\textit{c}) ratio of streamwise to vertical fluctuations  (\textit{d}) Reynolds shear stresses $\langle uv \rangle/U_1^2$.}
\label{fig:turb_contours}
\end{figure}

Streamwise and vertical turbulent fluctuations  reached up to 50 $\%$ of the inflow velocity  at a  region next to the jump toe (figures \ref{fig:turb_contours} (\textit{a},\textit{b})), exceeding common turbulence quantities of wall-jet flows. The velocity fluctuations were possibly generated by large pressure fluctuations throughout a wedge-shaped region generated by the jump toe. This region was visually characterised by droplet ejection processes and large splashing, 
closely following the aeration theories of \cite{Brocchini01JFM} and \cite{Valero18Reformulating}. Similar to plane wall-jets, streamwise fluctuations were larger than vertical fluctuations in the region next to the wall (figure \ref{fig:turb_contours} (\textit{c})). However, vertical fluctuations exceeded streamwise fluctuations in the upper part of the roller (figure \ref{fig:turb_contours} (\textit{c})), showing clear differences between the hydraulic jump and the wall-jet. These differences were caused by a recirculating air-water mixture, involving a highly dynamic free-surface with significant vertical movement above $y/y_{1/2}>1$ (see also figure \ref{fig:OFVelocities} (\textit{b})).

The Reynolds shear stresses  are shown in figure \ref{fig:turb_contours} (\textit{d}). $\langle u  v \rangle/U_1^2$ ranged from $-$0.05 to 0, with peaks occurring at $y = d_1$ and close to the free-surface at the impingement. Areas of null shear stresses were present inside the roller body, suggesting uncorrelated kinematics of the falling particles, similar to an inviscid flow. Downstream of the roller, the free-surface was readily identified and exhibited a wavy pattern. The flow was characterised by low void fractions and lower turbulence levels when compared to the impinging region. 
 
\begin{figure}[h!]
\centerline{\includegraphics[width=\textwidth]{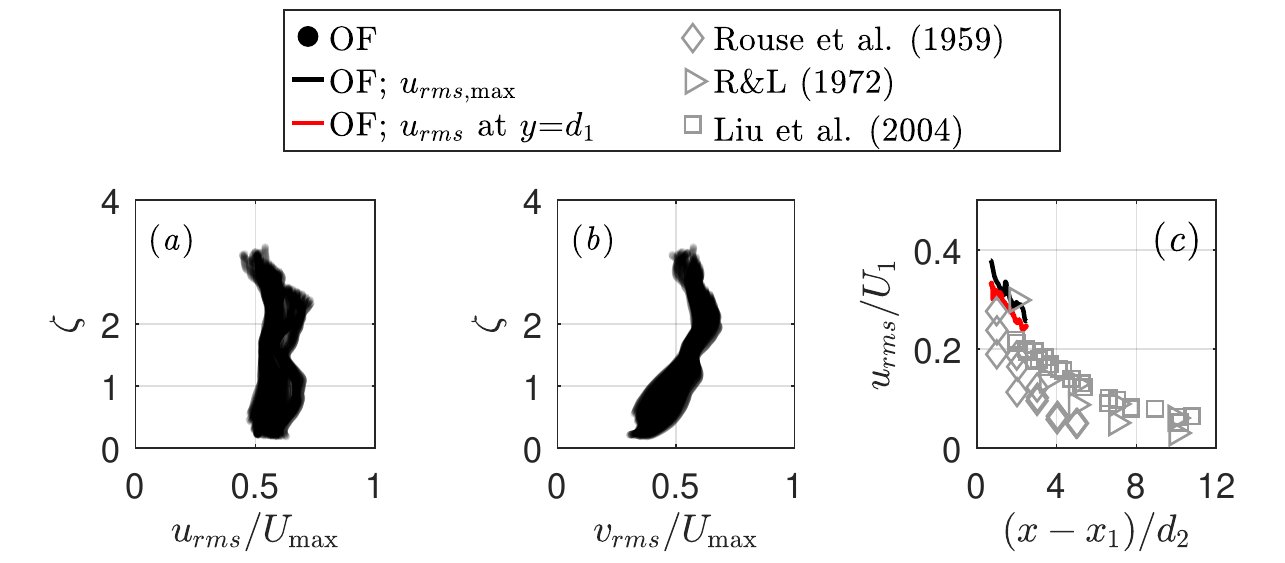}}
\caption{Self-similarity of turbulent OF velocity fluctuations - comparison with monophase data from \cite{Rouse58}, \cite{Resch72} and \cite{Liu04} (\textit{a}) streamwise velocity fluctuations for $\mathcal{X} > 8$ \protect (\textit{b}) vertical velocity fluctuations for $\mathcal{X} > 8$ \protect (\textit{c}) maximum cross-sectional fluctuations.} 
  \label{fig:turb_profiles}
\end{figure}

The distribution of velocity fluctuations (figures \ref{fig:turb_profiles} (\textit{a},\textit{b})) was qualitatively in accordance with monophase acoustic Doppler velocimeter (ADV) experiments of \cite{Liu04}. Maximum fluctuations occurred in the upper region of the jump, following an exponential decay in the streamwise direction. A detailed comparison with monophase measurements revealed that maximum two-phase OF velocity fluctuations were greater than previously reported (figure \ref{fig:turb_profiles} (\textit{c})). There are several possible explanations: 1.) image-based methods allow access to the complete velocity field and are not constrained to regions with less aeration. ADV methods did not measure within the most turbulent jump toe region 2.) fluctuations generated in the impingement region may scale with the Froude number and 3.) turbulence modulation becomes more relevant with larger aeration \citep{Balachandar10}.
Similar findings were also published in \cite{Lin12}. For future reference, velocity fluctuations at $y = d_1$ were included in figure \ref{fig:turb_profiles} (\textit{c}).

\subsection{Turbulent integral scales}
Integral time and length scales were obtained through the corresponding auto-correlation functions. For the streamwise velocity fluctuations, the time auto-correlation function can be defined as \citep{Pope20}:

\begin{equation}
S_{uu}( \tau) = \langle u(t) \, u(t+ \tau) \rangle
\label{eq:Suu}
\end{equation}
with $\tau$ being the lag time. Similarly, the spatial auto-correlation is defined as: % \citep{Pope20}:

\begin{equation}
R_{uu}(r_1) = \langle u(x) \, u(x+r_1) \rangle
\label{eq:Ruu}
\end{equation}
$r_1$ being the streamwise lag distance. %from a point at coordinate $x$. 
Equation (\ref{eq:Suu}) describes the dependence of the streamwise fluctuating velocity component at time $t$ on the value at a lagged time $t+\tau$, which is likewise for the spatial component (eq. (\ref{eq:Ruu})). Therefore, the auto-correlation function is a measure of similitude and the associated scales represent the memory of the described process.

The auto-correlation functions were numerically integrated to obtain integral time and length scales in the streamwise direction:

\begin{eqnarray}
T_{uu,x} = \int _0 ^\infty  S_{uu}(\tau) \, \mathrm{d} \tau\\
L_{uu,x} = \int _0 ^\infty  R_{uu}(r_1) \, \mathrm{d} r_1 \label{Luu}
\end{eqnarray}
In practice, numerical integration was performed up to the first zero-crossing. Integration of eq. (\ref{Luu}) accounted for both auto-correlations (positive and negative $x$-directions) when available within the studied domain. 

\begin{figure}[h!]
\centerline{\includegraphics[width=\textwidth]{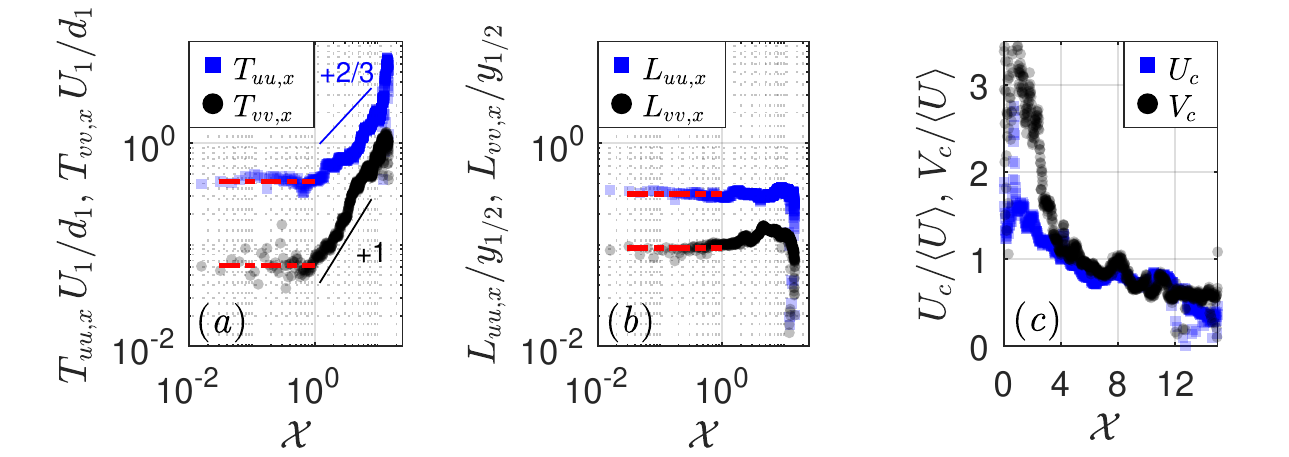}}
\caption{Turbulent scales at $y = d_1$. (\textit{a}) Integral time scales $T_{uu,x}$ and $T_{vv,x}$ (\textit{b}) integral length scales $L_{uu,x}$ and $L_{x,vv}$ (\textit{c}) velocity scales $U_c$ and $V_c$.}
\label{fig:TuuLuu}
\end{figure}

Figure \ref{fig:TuuLuu} shows the turbulent integral time, length scales and velocity scales at $y = d_1$. The integral time scales, dimensionless using $U_1$, remained constant within the impinging region (\mbox{$\mathcal{X}<$ 1}) at values of $T_{uu,x} U_{1}$ $\approx$ 0.42 and $T_{uu,x} U_{1}/d_{1}$ $\approx$ 0.06 (figure \ref{fig:TuuLuu} (\textit{a})). In the flow establishment zone ($1<\mathcal{X}<$ 5), increasing time scales indicated a deceleration of vortices. The growth rate of  $T_{vv,x}$ was larger than the growth rate of $T_{uu,x}$, hinting a transition to isotropic turbulence. Downstream of \mbox{$\mathcal{X} \approx$ 5}, the growth rates of $T_{uu,x}$ and $T_{vv,x}$ were similar and their ratio was approximately constant at $T_{uu,x}/T_{vv,x}$  $\approx$ 2.6.

Dimensionless length scales remained constant next to the impingement region with values of $L_{uu,x} / y_{1/2} \approx$ 0.32 and $L_{vv,x} / y_{1/2} \approx$ 0.09 (figure \ref{fig:TuuLuu} (\textit{b})). For $\mathcal{X}>1$, vortices grew with rates very similar to the spreading rate of the jump. The ratio of streamwise and vertical length scales was $L_{uu,x}/L_{vv,x}$ $\approx$ 2.7 for \mbox{$\mathcal{X}<$ 12}, which was almost identical to the ratio of the time scales.

The velocity scale $U_c = L_{uu,x}/T_{uu,x}$ varied linearly from $U_c/\langle U \rangle$ = 1.7 to 0.8 between $\mathcal{X}=$ 0 and 5 (figure \ref{fig:TuuLuu} (\textit{c})). Downstream of the flow establishment zone, the velocity scale remained roughly at $U_c/\langle U \rangle$= 0.8  up to $\mathcal{X} \approx$ 12. The vertical velocity scale $V_c = L_{vv,x}/T_{vv,x}$ behaved similar to the streamwise velocity scale, but had a considerably larger peak next to the jump toe.

\subsection{Velocity spectra}
The velocity fluctuation spectrum represents the distribution of energy by frequency and is defined as \citep{Pope20}:

\begin{equation}
  E_{u u} (f) =
  \frac{2}{\pi}
  u_{rms}
   ^2
  \int_0 ^{\infty} 
  S_{u u} (\tau) \cos (f \, \tau) \,
  \mathrm{d} \tau
\label{eq:spectra_definition}
\end{equation}

\begin{figure}[h!]
\centerline{\includegraphics[width=\textwidth]{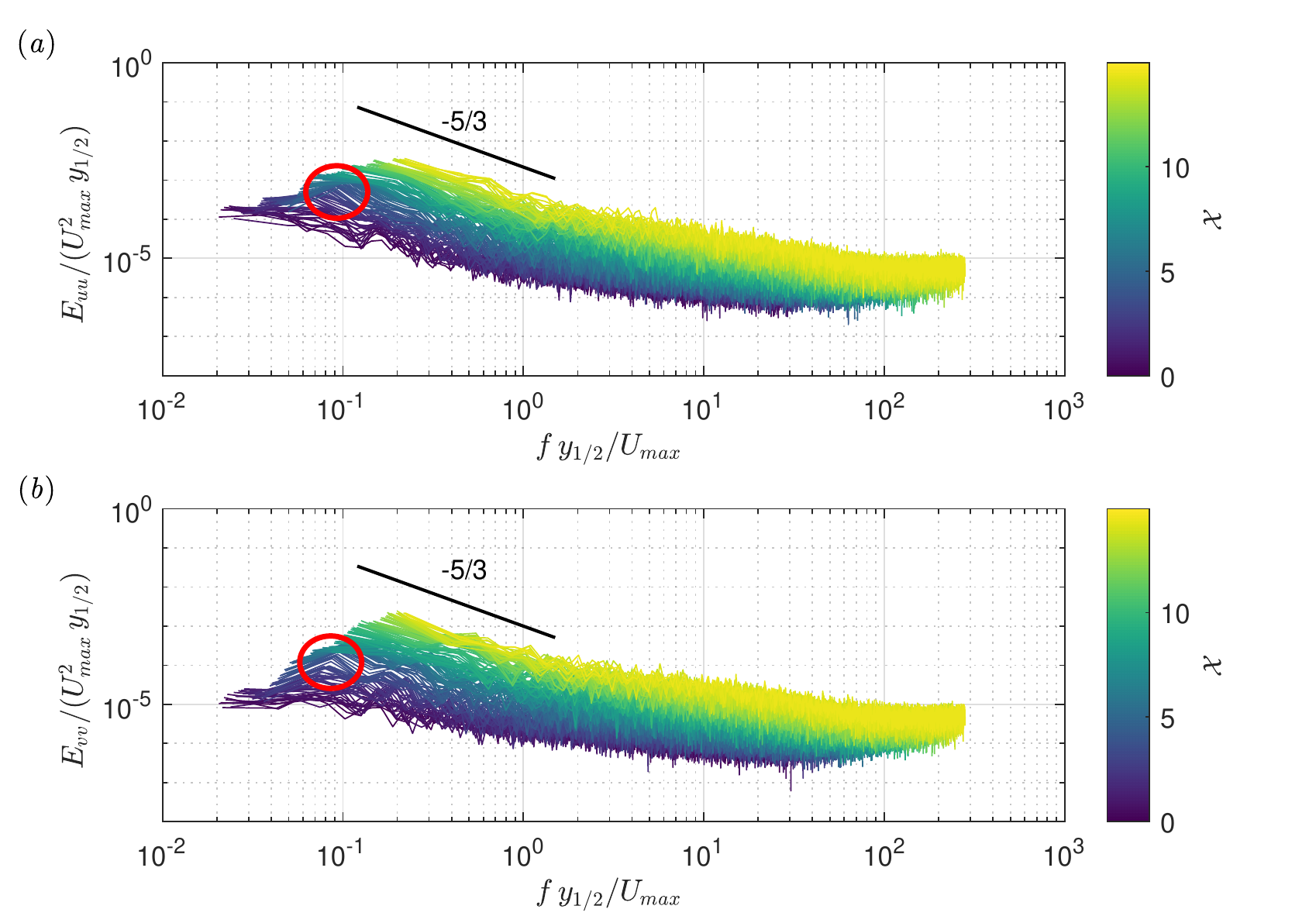}}
\caption{Velocity spectra at $y = d_1$; for clarity,  spectra are plotted every 5th pixel in streamwise direction; red circles indicate observed frequency peaks (\textit{a}) streamwise velocity spectra $E_{uu}$ (\textit{b}) vertical velocity spectra $E_{vv}$.}
\label{fig:EspectraAll}
\end{figure}

Spectra were computed at $y = d_1$ using fast Fourier transformations based on the \cite{Welch67} method, with windows of 1/5$^\mathrm{th}$ of the sampling time and 50 $\%$ overlap. The most energetic spectra (absolute values) were found next to the jump toe. In dimensionless terms, the streamwise and vertical velocity spectra collapsed approximately at \mbox{5 $<\mathcal{X}<$ 8} (figure \ref{fig:EspectraAll}).

In both streamwise and vertical velocity spectra, a $-$5/3 slope was satisfied for the inertial range, but a subtle change was noticed at higher frequencies. One possible explanation for the flattening of the slope could be related to the bubble-eddy interaction, transferring energy from lower frequencies to those in the scale of the bubbles. This turbulence modulation phenomena would imply an increase of the turbulence energy at large frequencies.

Along the flow establishment zone ($1<\mathcal{X}<5$), clear frequency peaks were observed at Strouhal numbers  $St = f \, y_{1/2}/U_{max} \approx$ 0.10 (indicated with red circles in figure \ref{fig:EspectraAll}), relating to a frequency of 4 Hz. This frequency was slightly larger than free-surface fluctuation frequencies for similar flow conditions \citep{Montano18}. As the frequency of \mbox{4 Hz} was not present throughout the impinging region, it becomes clear that this is a consequence of vortex shedding at the downstream end of the jump toe, recirculated in the roller. Because of this recirculation and a damping effect within the roller, it is expected that free-surface frequencies are smaller than vortex shedding frequencies. Consequently, free-surface frequencies would reduce with longer rollers and larger inflow Froude numbers, as previously observed in \cite{Montano18}.

\section{Conclusion} \label{sec:conclusions}
This study investigated a stable hydraulic jump (with \mbox{$Fr_1$ = 4.25} and \mbox{$Re$ = 1.15$\times$10$^5$}) using recent experimental techniques for air-water flows, including phase-detection intrusive probes (point measurements) and image-based velocimetry (restricted to side-channel view). Basic flow parameters  were compared to previous literature and to analytical-empirical models, confirming the potential of the optical flow technique. For the first time, measurements were expanded to incorporate fluctuating velocity components for a stable and highly-aerated hydraulic jump.

Based on the present observations, the impingement region and the flow establishment zone were defined for $\mathcal{X} <$ 1 and $1 < \mathcal{X}< 5$, followed by self-similar profiles in terms of mean flow quantities and turbulence intensities. Computed velocity spectra showed a dominant vortex shedding frequency of 4 Hz, originating at the downstream end of the jump toe. The velocity spectra collapsed in the roller region and exhibited a $-$5/3 slope in the inertial subrange, although some deviations were observed for high frequencies. This could be linked to a transfer of energy from the inertial range to frequencies associated with the scale of bubbles.

Overall, this study sheds light on turbulent processes  in the most energetic region of a stable hydraulic jump, providing validation data for high-fidelity numerical models. A meaningful extension of this work would involve the application of the current methodology over a wider range of Froude and Reynolds numbers, as well as the investigation of fully developed versus partially developed inflow conditions.

\section*{Acknowledgements}
The authors thank The University of Queensland’s School of Civil Engineering, which provided full-access to the UQ AEB Hydraulics laboratory. Jason Van Der Gevel and Stewart Matthews (The University of Queensland) are thanked for their technical assistance. The discussions with Hubert Chanson are acknowledged. Ana Gencic is thanked for proof-reading the manuscript.

\section*{Funding}
 Matthias Kramer was supported by the German Research Foundation [grant number KR 4872/2-1].
 
\section*{Notation} \label{sec:notation}
\vspace{-0.5cm}
\begin{tabbing}
\hspace*{0cm}\=\hspace*{1.5cm}\=\hspace*{10.5cm}\=\kill\\
\>$\textbf{A}$  \> symmetric matrix for OF calculations \> (-)\\
\>$\textbf{b}$  \> vector for OF calculations \> (-)\\
\>$b_o$  \> empirical coefficient for evaluating the roller length \> (-)\\
\>$C$  \> void fraction \> (-)\\
\>$D^*$  \> dimensionless diffusivity of the recirculation region \> (-) \\
\>$D^\#$  \> dimensionless diffusivity of the shear region \> (-)\\
\>$d$  \> free-surface elevation\> (m)\\
\>$d_1$  \> mean flow depth at the inflow section \> (m)\\
\>$d_2$  \> conjugate mean flow depth \> (m)\\
\>$E_{uu}$ \> streamwise velocity fluctuation spectrum \> (m$^3$/s$^2$)\\
\>$E_{vv}$ \> vertical velocity fluctuation spectrum \> (m$^3$/s$^2$)\\
%v\>$\textbf{e}_1$  \> unit vector in the $x$-coordinate direction\> (-)\\
\> F  \> filter-size of the Farneback method \> (px)\\
\>$F$  \> bubble/droplet count rate \> (s$^{-1}$)\\
\>$Fr_1$  \> inlet Froude-number, defined as $Fr = U_1/\sqrt{gd_1}$ \> (-)\\
\>$g$  \> gravitational acceleration \> (m s$^{-2}$)\\
\> g  \> threshold of the image gradient magnitude \> (px$^{-1}$)\\
\>$I$  \> pixel brightness intensity \> (-)\\
\>$K$  \> integration constant \> (-)\\
\>$L_j$  \> length of the hydraulic jump \> (m)\\
\>$L_r$  \> roller length of the hydraulic jump \> (m)\\
\>$L_{uu,x}$ \> streamwise integral turbulence length scale in $x$-direction \> (s)\\
\>$L_{vv,x}$ \> vertical integral turbulence length scale in $x$-direction \> (s)\\
\> N  \> neighbourhood-size of the Farneback method \> (px)\\
\>$N$  \> power law constant \> (-)\\
\>$N_p$  \> number of carrier phase chords \> (-)\\
\>$q$  \> specific water discharge \> (m$^2$ s$^{-1}$)\\
\>$R_{uu}$ \> space auto-correlation function \> (-)\\
\>$Re$  \> Reynolds number, defined as $Re = 4q/\nu$ \> (-)\\
\>$R_{12}$  \> cross-correlation coefficient \> (-)\\
\>$r_1$  \> lag distance centred at coordinate $x$ \> (m)\\
\>$S$  \> spreading rate of the jump \> (-)\\
\>$S_1$  \> binarized leading tip signal \> (-)\\
\>$S_2$  \> binarized trailing tip signal \> (-)\\
\>$St$  \> Strouhal number, defined as $St =f \, y_{1/2}/ U_{max} $  \> (-)\\
\>$S_{uu}$ \> time auto-correlation function \> (-)\\
\>$T$  \> measurement interval \> (s)\\
\>$T_i$  \> time delay for AWCCT \> (s)\\
\>$T_{uu,x}$ \> streamwise integral turbulence time scale in $x$-direction \> (s)\\
\>$T_{vv,x}$ \> vertical integral turbulence time scale in $x$-direction \> (s)\\
\>$t$  \> time \> (s)\\
\>$U$  \> instantaneous streamwise velocity \> (m s$^{-1}$)\\  
\>$U_1$  \> depth-averaged streamwise velocity at $x$ = $x_1$ \> (m s$^{-1}$)\\  
\>$ U_{\mathrm{max}} $  \> maximum time-averaged cross-sectional velocity \> (m s$^{-1}$)\\
\>$ U_{c} $  \> velocity scale \> (m s$^{-1}$)\\ 
\>$ U_{\mathrm{recirc}} $  \> time-averaged recirculation velocity \> (m s$^{-1}$)\\             

\>$u$  \> fluctuating streamwise velocity \> (m s$^{-1}$)\\
 
\>$V$  \> instantaneous normal velocity \> (m s$^{-1}$)\\
\>$v$  \> fluctuating normal velocity \> (m s$^{-1}$)\\ 

\>$W$  \> channel width \> (m)\\
\>$\mathcal{W}_{T,i}$  \> time window for AWCCT \> (s)\\
\>$We$  \> Weber number, defined as $We = \rho U_1^2 d_1/\sigma$ \> (-)\\
 \>$\mathcal{X}$  \> dimensionless streamwise coordinate, $\mathcal{X} = (x-x_1)/d_1$  \> (-)\\                
\>$x$  \> streamwise coordinate \> (m)\\   
\>$\textbf{x}$  \> coordinate vector for OF calculations \> \\  
\>$x_1$  \> jump-toe position, related to the upstream gate \> (m)\\  
\>$y$  \> vertical coordinate \> (m)\\ 
\>$y_{1/2}$  \> half-width of the jump \> (m)\\
\>$y_{50}$  \> elevation where $\langle C \rangle$ = 0.5 \> (m)\\
\>$y*$  \> elevation with a local void fraction minimum \> (m)\\     
\>$y_{U_\mathrm{max}}$  \> elevation where $\langle U \rangle  = U_{\mathrm{max}}$\> (m)\\                   
\end{tabbing}
Greek letters
\vspace{-0.5cm}
\begin{tabbing}
\hspace*{0cm}\=\hspace*{1.5cm}\=\hspace*{10.5cm}\=\kill\\
\>$\alpha$  \> $\approx$ $\ln 2$ \> \\
\>$\Delta t_i $  \> travel time of interfaces  \> (s)\\
\>$\Delta x $  \> streamwise tip separation \> (m)\\
\>$\Delta \textbf{x} $  \> displacement vector \> \\
\>$\zeta$  \> similarity variable, defined as $\zeta = y/y_{1/2}$ \> \\
\>$\eta$  \> similarity variable \> \\
\>$\nu $  \> kinematic viscosity of water \> (m$^{2}$ s$^{-1}$)\\
\>$\nu_t$  \> turbulent viscosity \> (m$^{2}$ s$^{-1}$)\\
\>$\rho $  \> density of water \> (kg m$^{-3}$)\\
\>$\sigma $  \> surface tension \> (N m$^{-1}$)\\
\>$\tau $  \> time lag of the cross-correlation function \> (s)\\

\end{tabbing}
Operators and subscripts
\vspace{-0.5cm}
\begin{tabbing}
\hspace*{0cm}\=\hspace*{1.5cm}\=\hspace*{10.5cm}\=\kill\\
\>$\times$  \> multiplication\\
\>$\langle ... \rangle $  \> time-average over measurement time $T$\\
\>$rms$  \> square root of mean velocity fluctuations\> \\
\end{tabbing}

\bibliography{HJ}

\newpage
\appendix
\section{Velocity estimation techniques}
\subsection{Optical flow velocity estimation with the Farnebaeck method}
\label{Farnebackmethod}
The recorded image sequences allowed for the estimation of the two-dimensional optical flow (OF) velocity using the local \cite{Farneback2002,Farneback2003} method. The pixel intensity $I$ was approximated with quadratic polynomials as:

\begin{equation}
I_1(x,y) = \textbf{x}^T \textbf{A}_1 \textbf{x} + \textbf{b}^T_1 \textbf{x} + c_1
\end{equation}
where $\textbf{x}$ is a coordinate vector, $\textbf{A}_1$ is a symmetric matrix, $\textbf{b}_1$ is a vector, $c_1$ is a scalar and the index 1 refers to the first image of an image pair. The pixel intensity pattern of the second image can be constructed by taking a displacement $\Delta\textbf{x}$ into account:
{\small
\begin{eqnarray}
I_2(x,y) &= &I_1(\textbf{x} - \Delta\textbf{x}) =  (\textbf{x} - \Delta\textbf{x})^T \textbf{A}_1 (\textbf{x} - \Delta\textbf{x})  + \textbf{b}^T_1 (\textbf{x} - \Delta\textbf{x}) + c_1 \label{I}\\
&=& \textbf{x}^T \textbf{A}_1 \textbf{x} + (\textbf{b}_1-2 \textbf{A}_1 \Delta\textbf{x})^T \textbf{x} + \Delta\textbf{x}^T \textbf{A}_1 \Delta\textbf{x} - \textbf{b}_1^T \Delta\textbf{x} + c_1 \label{I1}\\
&=& \textbf{x}^T \textbf{A}_2 \textbf{x} + \textbf{b}^T_2 \textbf{x} + c_2 \label{I2}
\end{eqnarray}}
The conservation of the pixel intensity is tested by comparing the polynomial coefficients of Equations (\ref{I1}) and (\ref{I2}), yielding an expression for the displacement vector \citep{Farneback2003}:
\begin{eqnarray}
2 \textbf{A}_1 \Delta\textbf{x} = -(\textbf{b}_2 - \textbf{b}_1)\\
\Delta\textbf{x} = -\frac{1}{2} \textbf{A}_1^{-1}(\textbf{b}_2 - \textbf{b}_1)
\label{Eq6} 
\end{eqnarray}

Given that the pointwise solution of Equation (\ref{Eq6}) might be too noisy, the optical flow is commonly integrated over a specified neighbourhood-size, assuming that there is only little variation in the displacement field. 

The Farnebaeck method was previously used for optical flow velocity estimation in highly aerated spillway flows \citep{Zhang17,Kramer18OF}. \cite{Bung17} used synthetic particle images to show that this technique is suitable for turbulence analysis \citep{Bung17}. 
In the present study, optical flow velocities were estimated with a neighbourhood-size of N = 5 px, a filter-size of F = 15 px and an image pyramid with two levels, taking recent  sensitivity analyses into account \citep{Kramer18OF}. Note that one pixel on the image plane corresponded to a physical lenght of 0.66 mm. Additional filtering was performed to detect outliers. These filtering techniques included: 
\begin{itemize}
\item  Image-gradient filter (indicator function) for removing erroneous data and filtering foreground movement of air-water interfaces. Similar to \cite{Kramer18OF}, an image gradient magnitude threshold of g = 5 px$^{-1}$ was applied.
\item The velocity time series for each pixel were processed with the despiking method of \cite{Goring02}, as modifed by \cite{Wahl03}, without taking velocity gradients into consideration.
\end{itemize}

\begin{figure}[h!]
\centerline{\includegraphics[width=\textwidth]{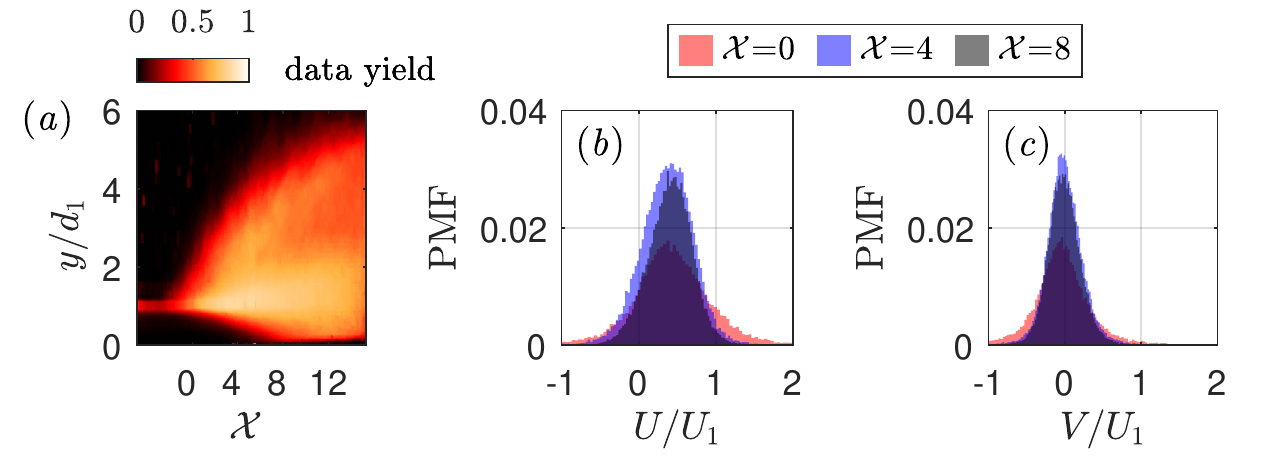}}
\caption{Results of the applied filtering methodology (\textit{a}) data yield (\textit{b}) probability mass functions of streamwise velocities at $\mathcal{X}$ = 0, 4 and 8 ($y = d_1$) (\textit{c}) probability mass functions ($y = d_1$) of vertical velocities at $\mathcal{X}$ = 0, 4 and 8 ($y = d_1$).}
\label{fig:datayield}
\end{figure} 

Figure \ref{fig:datayield} (\textit{a}) shows the data yield of the image-based velocimetry after filtering. 
The amount of valid data was highest in the shear layer and in the upper region of the roller, which was qualitatively in accordance with the distribution of the bubble/droplet count rate, confirming the selection of the image-gradient filter. Probability mass functions (PMFs) of streamwise and vertical velocities (shown for $y = d_1$) were similar to a Gaussian. A broad distribution was observed close to the impingement region (figures \ref{fig:datayield} (\textit{b}) and (\textit{c})), revealing jump toe fluctuations in combination with a highly dynamic free-surface. 

\begin{figure}[h!]
\centerline{\includegraphics[width=0.95\textwidth]{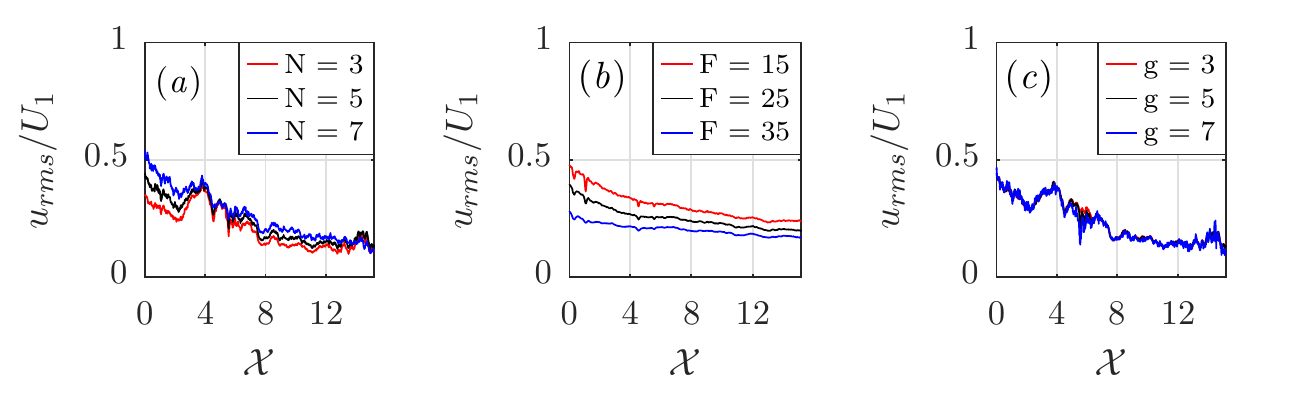}}
\caption{Sensitivity analysis of the Farnebaeck method for the streamwise velocity fluctuations at $y = d_1$; reference parameters: neighbourhood-size N = 5 px, filter-size F = 15 px, gradient magnitude threshold g = 5 px$^{-1}$ (\textit{a}) different filter-sizes (\textit{b}) different neighbourhood-sizes (\textit{c}) different gradient magnitude thresholds.}
\label{fig:sensitivity}
\end{figure}

A sensitivity analysis was performed to show the effect of the processing parameters of the Farneback method on estimated velocity fluctuations. The analysis was conducted with reference parameters N = 5 px, F = 15 px and g = 5 px$^{-1}$. A larger neighbourhood-size resulted in higher fluctuations (figure \ref{fig:sensitivity} (\textit{a})), whereas a larger filter-size yielded smaller fluctuations (figure \ref{fig:sensitivity} (\textit{b})). Possible reasons for such behaviour are 1.) the inclusion of image noise with increasing neighbourhood-size and 2.) the filtering is done after the computation of the displacement, consequently leading to a reduction of velocity fluctuations with larger filter-sizes. In contrast, the threshold selection of the image-gradient filter did not affect the estimated velocity fluctuations (figure \ref{fig:sensitivity} (\textit{c})). 

\subsection{AWCC technique for phase-detection probe signals}
\label{AWCCT}
The adaptive window cross-correlation (AWCC) technique is a processing method for dual-tip phase-detection probe signals, developed by \cite{Kramer19AWCC} and available under \cite{AWCC}. The technique relies upon adaptive time windows for cross-correlation analysis together with robust filtering criteria. 

The phase-detection probe signals were segmented into very short windows, based on a defined number of bubble-droplet events. A segment started when the dispersed phase was detected by the leading tip and finished after a number ($N_p$) of carrier phase chords. The time shift $\Delta\, t_i$ for an arbitrary window $\mathcal{W}_{T, i}$ was obtained through cross-correlation of the leading and trailing tip signals, $S_{1,i}$ and $S_{2,i}$
{\small
\begin{equation}
 R_{12, i}(\tau) 
  = 
 \frac{\sum_{t=t_i}^{t_i+\mathcal{W}_{T, i}} (S_{1}(t)-\langle S_{1,i} \rangle) \times (S_{2}(t+\tau)- \langle S_{2,i}\rangle)}
 { \sqrt{\sum_{t=t_i}^{t_i+\mathcal{W}_{T, i}} (S_{1}(t)-\langle S_{1,i}\rangle)^2} \times \sqrt{\sum_{t=t_i}^{t_i+\mathcal{W}_{T, i}}(S_{2}(t+\tau)-\langle S_{2,i} \rangle)^2 }} 
\label{eq:xcor_basic}
\end{equation}}
with $\tau$ being the time lag, $t_i$ the starting time step of the segment $i$ and $R_{12, i}$ the correlation coefficient. The peak of $R_{12,i}$ indicated the time delay $T_i =\arg max(R_{12, i})$ for which both signals were best correlated, allowing to approximate $\Delta\, t_i \approx T_i$. Hence, a longitudinal pseudo-instantaneous velocity, representative of the window time $ \mathcal{W}_{T,i}$, was computed as
 
\begin{equation}
\left[ \, \langle U \rangle \, \right]_{t_i} ^{t_i + \mathcal{W}_{T,i}}
\approx U_i = \frac{\Delta\, x}{T_i}
\end{equation}
with $\Delta\, x$ being the streamwise separation distance of the two probe tips. To ensure robust velocity estimation, two filtering criteria and an outlier detection method were applied. The filtering criteria implied 1.) a minimum similarity between both segments, expressed through a threshold of the required maximum  cross-correlation coefficient $R_{12, i, \mathrm{max}}$ and 2.) a secondary peak ratio SPR, defined as the ratio of the second tallest peak to the first tallest peak of the cross-correlation function. The despiking method of \cite{Goring02}, as modifed by \cite{Wahl03}, was used to reduce the number of outliers without taking velocity derivatives into account.

The flow in hydraulic jumps is characterised by a highly 3-dimensional motion of air and water, especially close to the toe and in the upper roller region. Because of these challenging flow conditions, the filtering thresholds of the AWCC technique were set to values of  $R_{12, i, \mathrm{max}} > 0.3$ and  $\mathrm{SPR}_i < 0.7$. The number of particles was chosen following a sensitivity analysis to \mbox{$N_p$ = 15}. 

\begin{figure}[h!]
\centerline{\includegraphics[width=0.9\textwidth]{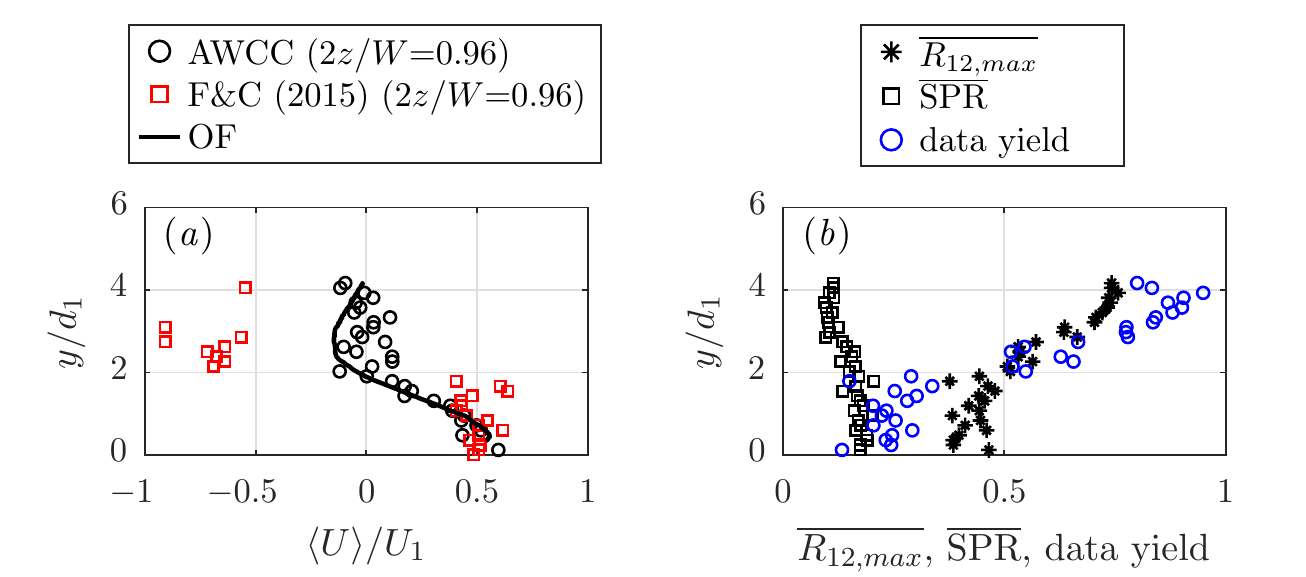}}
\caption{Vertical profile measured at $\mathcal{X}=7.1$; hydraulic jump with $Fr_1$ = 4.25 and $Re$ = 1.15 $\times$ 10$^4$ (\textit{a}) Comparison of mean velocities computed with OF, AWCC and signal processing after \cite{Felder2015} (\textit{b}) AWCC parameters, including maximum cross-correlation coefficient ($\overline{R_{12,\mathrm{max}}}$), secondary peak ratio ($\overline{\textrm{SPR}}$) and data yield. Processing parameters were $N_p$ = 15; SPR = 0.7; $R_{12,max}$ = 0.3.}
\label{fig:AWCC}
\end{figure}

Because of very low void fractions and low bubble/droplet count rates (figure \ref{fig:phase-detection}), the measurement of velocity time series and turbulent quantities was not posible in proximity to the sidewall (at 2$z/W$=0.96). However, the AWCC technique allowed for a reliable characterisation of the jump's shear layer. Figure \ref{fig:AWCC} (\textit{a}) shows a comparison of mean streamwise velocities obtained with optical flow (OF) and phase-detection conductivity probe (CP), using 1.) AWCC and 2.) conventional signal processing after \cite{Felder2015} as applied, amongst others, by \cite{Chanson09,Chanson11,Wang14,Wang19}. The conventional approach seemed to overestimate the recirculation velocity, whereas a good agreement between AWCC and OF was seen. The measurement quality was evaluated based on the data yield, $\overline{R_{12, \mathrm{max}}}$ and  $\overline{\mathrm{SPR}}$ (evaluted as mean values), indicating an increasing performance of the AWCC technique with distance from the bottom of the channel (figure \ref{fig:AWCC} (\textit{b})).

\end{document}